\documentclass[aps,prl,twocolumn,superscriptaddress,amsmath]{revtex4}

\usepackage{graphicx}

\begin{document}


\title{Universal Behavior of Extreme Price Movements in Stock Markets}



\author{Miguel A. Fuentes}
\email{fuentesm@santafe.edu}
\affiliation{Santa Fe Institute, Santa Fe, New Mexico, United States of America}
\affiliation{Center for Advanced Studies in Ecology and Biodiversity, Facultad de Ciencias Biol\'{o}gicas, Pontificia Universidad Cat\'{o}lica de Chile, Santiago, Chile}
\affiliation{Statistical and Interdisciplinary Physics Group, Centro At\'omico Bariloche, Instituto Balseiro and Consejo Nacional de Investigaciones Cient\'ificas y T\'ecnicas, Bariloche, Argentina}

\author{Austin Gerig}
\email{gerig@santafe.edu}
\affiliation{School of Finance and Economics, University of Technology, Sydney, Australia}
\affiliation{Santa Fe Institute, Santa Fe, New Mexico, United States of America}

\author{Javier Vicente}
\email{jvicente@santafe.edu}
\affiliation{Santa Fe Institute, Santa Fe, New Mexico, United States of America}


\date{\today}

\begin{abstract}
Many studies assume stock prices follow a random process known as geometric Brownian motion.  Although approximately correct, this model fails to explain the frequent occurrence of extreme price movements, such as stock market crashes.  Using a large collection of data from three different stock markets, we present evidence that a modification to the random model -- adding a slow, but significant, fluctuation to the standard deviation of the process -- accurately explains the probability of different-sized price changes, including the relative high frequency of extreme movements.  Furthermore, we show that this process is similar across stocks so that their price fluctuations can be characterized by a single curve.  Because the behavior of price fluctuations is rooted in the characteristics of volatility, we expect our results to bring increased interest to stochastic volatility models, and especially to those that can produce the properties of volatility reported here.
\end{abstract}



\maketitle


\section{Introduction}
The first theoretical study of stock prices modeled price differences as a simple random process -- now commonly known as a drunkard's walk \cite{Bachelier00}.  Although pioneering for its time, several modifications to this model have been needed.  First was the realization that prices move in relative amounts rather than absolute amounts, and that returns rather than price differences should be modeled as a random process\cite{Osborne59}.  Next, several papers showed that returns could not be described by a static Gaussian process because the tails of the return distribution are too fat, i.e., large price fluctuations occur much too frequently\cite{Mandelbrot63, Fama65}.  Numerous studies have tried to characterize and explain this phenomenon\cite{Mantegna95, BouchaudPotters03, Gabaix03, Viswanathan03, Farmer04, Bassler07}.  This is because understanding the probability of large returns is very important for asset allocation, option pricing, and risk management.  In spite of this work, there is still no accepted theoretical explanation for this feature\cite{Stanley03}.  Here we present evidence that the non-Gaussian, fat-tailed shape of the return distribution is explained by modeling returns as a random process with a slowly fluctuating standard deviation (or volatility).  Previously, we have found that this model works well for several stocks traded on the London Stock Exchange (e-print arXiv:0906.3841).  Here we test the model using a larger collection of stocks from different exchanges and different time periods.  We show that the return distribution for these stocks is similar in shape and well-fit by the model, and we present evidence that the tail of the distribution for each stock is determined by the properties of volatility for that stock.

The idea that volatility fluctuations cause non-Gaussian returns is not new -- it was originally suggested several decades ago and is known as the \emph{mixture-of-distributions} hypothesis\cite{Fama65, Press67, Praetz72, Clark73, Blattberg74}.  This hypothesis can explain the non-Gaussian shape of the return distribution, but it is unable to explain the apparent stability of the distribution over longer time scales.  To account for this stability, others have suggested what is known as the \emph{stable Paretian} hypothesis -- that returns are drawn unconditionally from a fat-tailed, stable distribution\cite{Mandelbrot63, Mantegna95, Lux96}.  Our model captures both the non-Gaussian shape and the apparent stability of the return distribution by assuming that volatility fluctuations are significant over long time scales but relatively small over short time scales.  The model can be summarized as follows:  On any single day, returns are well described by a Gaussian distribution.  Across days, weeks, and months, however, slow but significant fluctuations in volatility produce returns with different standard deviations.  When collecting returns from each of these periods into one plot, the return distribution no longer looks Gaussian, but is fat-tailed.  The distribution keeps this shape when aggregating returns over longer time scales because volatility is slowly varying.  Because this process occurs in a similar way across stocks, the distribution of returns for different stocks collapse onto one curve.

The results we present are produced using a large amount of data (of the order of $10^7$ data points) from three stock markets over three time periods: the London Stock Exchange (LSE) from May 2, 2000 to December 31, 2002, the New York Stock Exchange (NYSE) from January 2, 2001 to December 31, 2002, and the Spanish Stock Exchange (SSE) from January 2, 2004 to December 29, 2006.  These time periods partially overlap for the NYSE and LSE data and are different for the SSE data.  The time discrepancies are due to obtaining data from different sources, and the results we present appear robust over these differences.  For each market, we study two highly traded stocks that are from different market sectors: AstraZeneca (AZN) and Vodafone (VOD) from the LSE, International Business Machines (IBM) and General Motors (GM) from the NYSE, and Telef\'{o}nica (TEF) and Banco Santander (SAN) from the SSE.  We consider the electronic markets for these stocks during normal trading hours, and we measure returns whenever the mid-price of a stock fluctuates.  This approach allows us to study returns on the finest possible time scale.  When aggregating returns over longer time scales, we use non-overlapping intervals.  We measure price fluctuations, or returns, in the standard way\cite{BouchaudPotters03} as $r_{t}(\tau)=\ln p_{t+\tau} -\ln p_t$, where $p$ is the mid-price, $t$ is the time (which we update by one unit whenever the price changes), and $\tau$ is the time increment.  Because time is updated whenever the price changes, it is a measure of the number of events that have occurred and not a measure of `calendar' or `clock' increments.

\section{Analysis}
To model the features of the return distribution, we use a general approach that assumes a Gaussian process for its dynamics.  The probability distribution of returns is therefore\cite{Gardiner04}
\begin{equation}
p(r,\tau|\beta) = \sqrt{ \frac{\beta}{2\pi \tau} } \exp{\left(-\frac{\beta r^2}{2 \tau}\right)}.
\end{equation}
This is coupled with a slow variation of the inverse variance $\beta\equiv1/\sigma^2$, where $\sigma$ is the volatility, $\left\langle r^2(\tau)\right\rangle=\sigma^2\tau$.  By slow variation, we mean that $\beta$ fluctuations are negligible compared to price fluctuations when observed over the time scales we study here -- up to one trading day.  This is not inconsistent with shocks to volatility as long as those shocks are relatively infrequent.  Others have reported systematic fluctuations in intraday volatility (see \cite{Andersen97} and references within), but these fluctuations closely mimic trading activity within the day.  Because we probe returns over a fixed number of return causing events, fluctuations in trading activity are removed from the analysis.

$\beta$ fluctuations over time scales longer than one day can be characterized by a probability distribution $g(\beta)$.  Several papers have stated different functional forms for the distribution of volatility\cite{Praetz72, Clark73, BouchaudPotters03}.  We propose -- and the evidence presented here supports our assumption -- that $g(\cdot)$ is similar across stocks and close to a gamma distribution
\begin{equation}
g_{n,\beta_0}(\beta)=\frac{1}{\Gamma \left( \frac{n}{2} \right)} \left( \frac{n}{2 \beta_0} \right)^{\frac{n}{2}} \beta^{\frac{n}{2}-1} \exp \left(-\frac{n \beta}{2 \beta_0} \right).
\end{equation}
There are several simple explanations for why the inverse variance might have this distribution\cite{Bouchaud01, PlatenHeath06}.

\begin{figure}[t]
\centering
\includegraphics[width=3.4in]{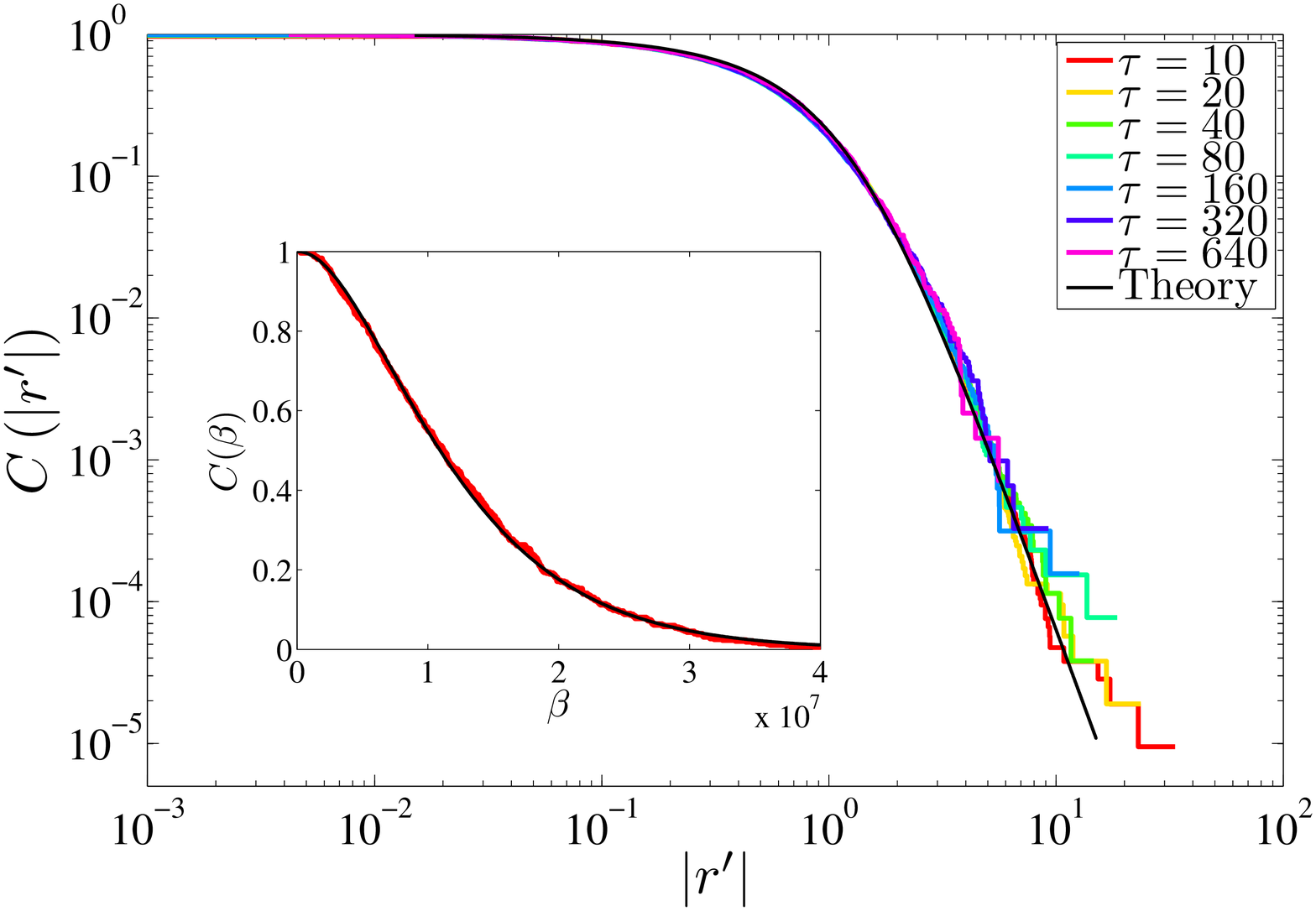}
\caption{{\bf Collapse of the complementary cumulative distribution (ccd) of absolute scaled returns, $C(|r'|)$, for the stock IBM.}  The ccd is shown for times scales $\tau=10$ to $\tau=640$.  The solid black line is the theoretical ccd using $\beta_0=1.28\times10^7$ and $n=4.40$ from fitting $\beta$ to a gamma distribution. Inset: ccd of the slow fluctuating variable $\beta$ for IBM, the red curve is the empirical ccd and the solid black line is a fit to a gamma distribution.}
\end{figure}

A straightforward integration of the conditional probability of returns, $p(r,\tau|\beta)$, and the distribution $g(\beta)$ yields the following for the return distribution:
\begin{equation}
P(r,\tau)=\frac{\Gamma \left[ \frac{(n+1)}{2}\right] }{\Gamma \left[ 
\frac{n}{2}\right] }\sqrt{\frac{\beta _{0}}{\pi n \tau}} \left( 1+
\frac{\beta _{0} r^{2}}{n\tau}\right) ^{-\frac{n+1}{2}},
\label{eq.probdist}
\end{equation}
which is a variant of the Student's $t$-distribution.  The non-Gaussian shape of the distribution results from collecting returns from time periods separated by long intervals where $\beta$ is different.  The stability of this shape for short to intermediate $\tau$ results from negligible fluctuations of $\beta$ over these time scales.

Although it is known that a gamma distributed inverse variance leads to a Student's $t$-distribution for returns\cite{Praetz72, Blattberg74}, this result does not explain how the return distribution retains its non-Gaussian shape for longer time scales.  To explain the persistence of the non-Gaussian shape, others have suggested that returns follow a fat-tailed stable distribution\cite{Mandelbrot63, Mantegna95, Lux96}.  In Eq.~\ref{eq.probdist}, we address both the non-Gaussian shape and the apparent stability of the return distribution -- both result from the properties of volatility that we have assumed in our model.

Other papers have reported that returns follow a Student's $t$-distribution and have fitted returns to a generic version of this distribution (see \cite{Praetz72, Blattberg74, BouchaudPotters03} for examples).  In the results we present below, we do not fit a Student's $t$-distribution, but instead compare the empirical distribution to the predicted distribution as expressed in Eq.~\ref{eq.probdist} and as determined by the independent measurement of $\beta_0$ and $n$.  This specifically tests the model rather than the more general result that returns follow a Student's $t$-distribution.

\section{Results}
In Fig.~1, we show the time collapse of the complementary cumulative distribution (ccd) of absolute scaled returns, $C(|r'|)$ with $r'=r\sqrt{2\beta_0/(n\tau)}$, for the stock IBM (the ccd is the integral of the probability function).  The ccd is plotted for $\tau=10$ to $\tau=640$, which is up to one trading day for the stocks in our study.  We show this plot in logarithmic coordinates to focus on the tails of the distribution, and we overlay the plot with the ccd of the theoretical distribution from Eq.~\ref{eq.probdist}.  As seen, the model matches the data well and the shape of the distribution is stable over these time scales.  The parameters $\beta_{0}$ and $n$ are determined using a maximum likelihood fit of $\beta$ to a gamma distribution, where $\beta$ is measured once per day.  In the inset of this figure, we show the ccd of $\beta$ compared to the fit.  Although not shown, these plots are very similar for the other stocks in our study.

\begin{figure}[t]
\centering
\includegraphics[width=3.4in]{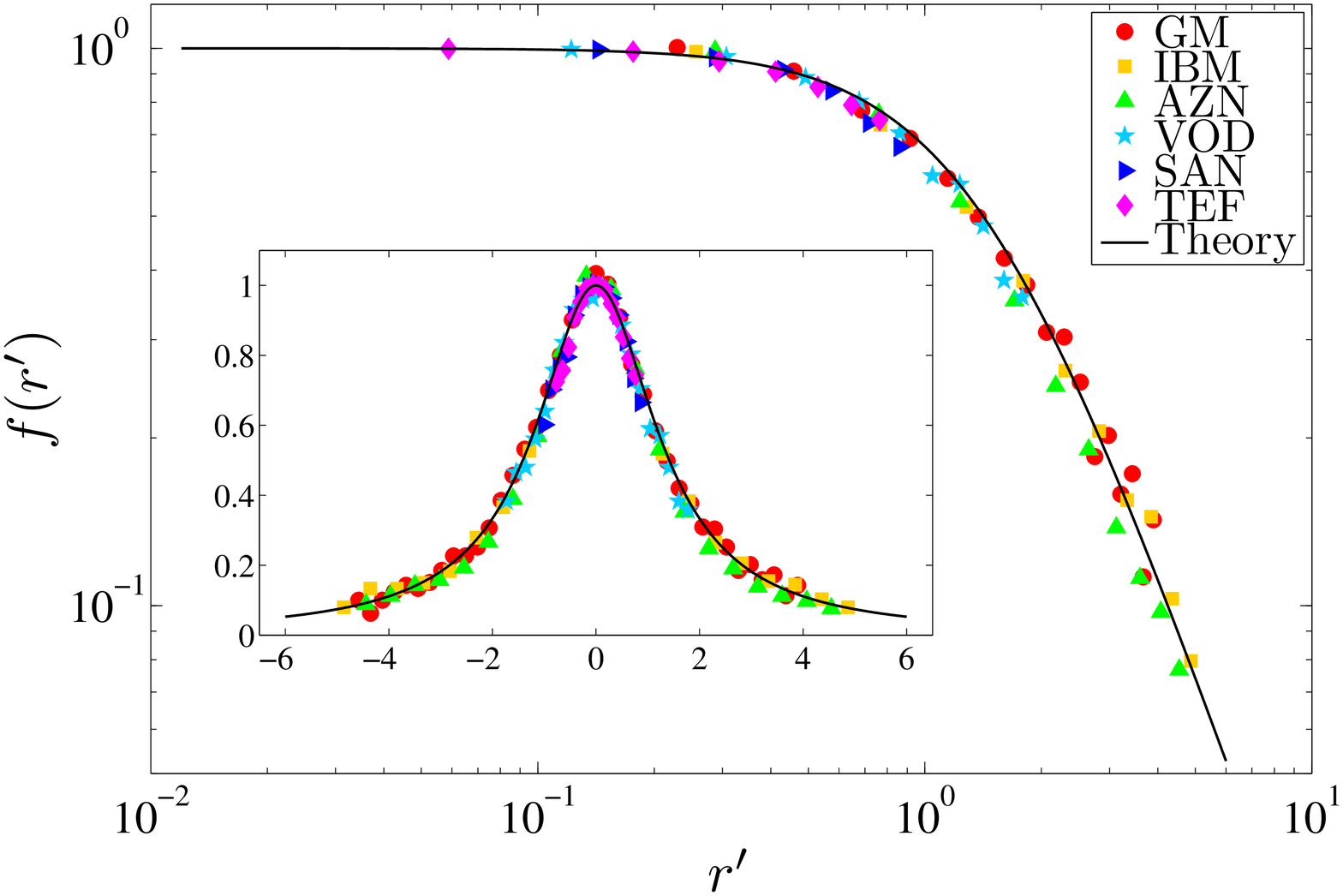}
\caption{{\bf Collapse of the return distribution on the function $f(r')$, Eq. 4, for the stocks in our study.} For each stock, the return distribution for $\tau=80$ is shown in logarithmic coordinates.  Inset: The same plot in regular coordinates.}
\end{figure}

\begin{figure}[t]
\centering
\includegraphics[width=3.4in]{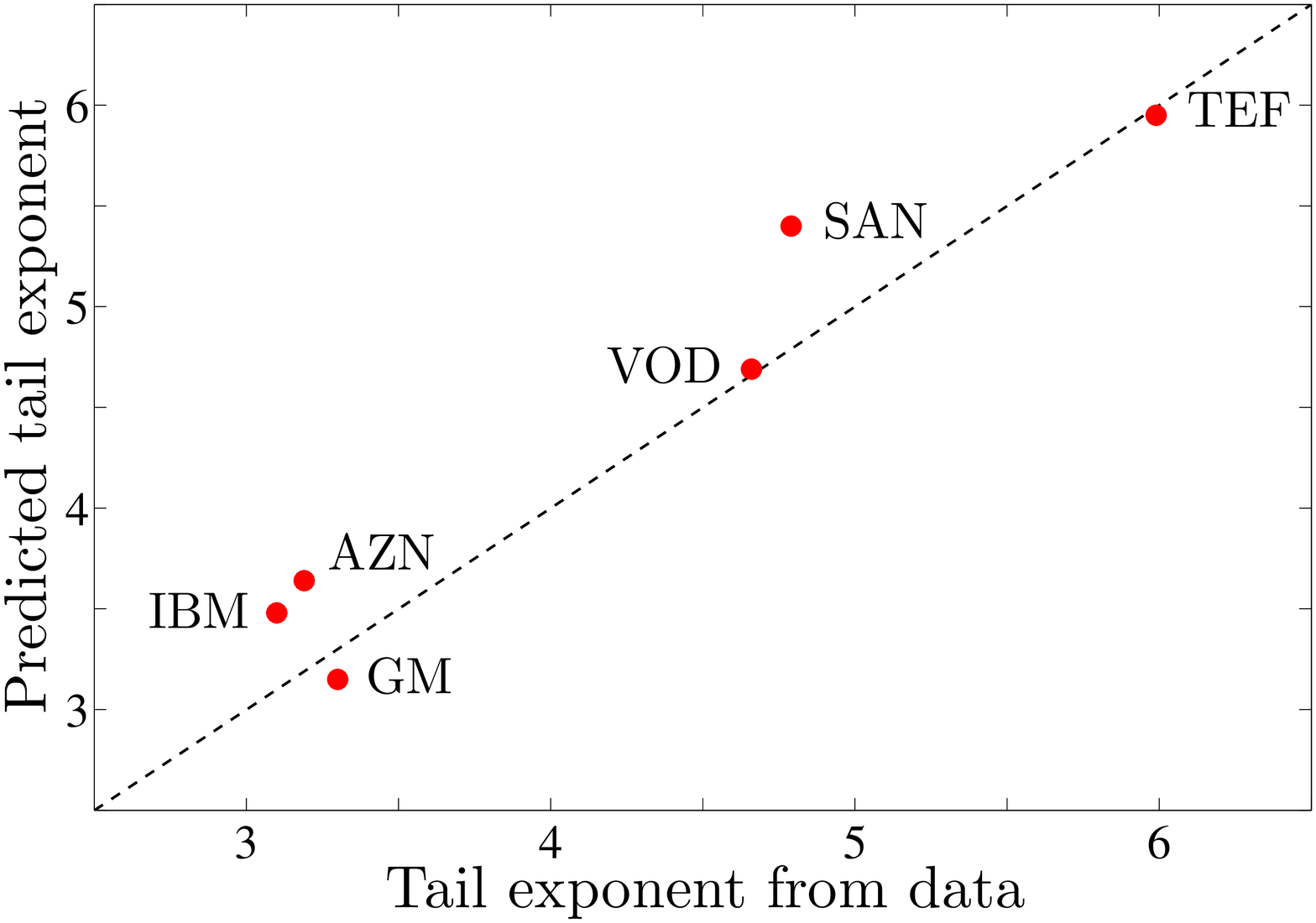}
\caption{{\bf Predicted vs. empirical tail exponent for the stocks under study.} The tail exponent is the asymptotic slope of the tail of the ccd when measured in logarithmic coordinates.  The dashed line shows $y=x$ for comparison only.}
\end{figure}

The above model assumes that the functional form of the return distribution is similar across stocks, and that differences are due to the particular properties of volatility for each stock.  This is verified in Fig.~2, where we show the collapse for {\it all} stocks using the following functional transformation, derived from the analytical results presented above:
\begin{equation}
f(r')= \left[\Lambda P(r',\tau)  \right]^{\frac{2}{n+1}}, 
\end{equation}
where $\Lambda=\sqrt{2\pi} \ \Gamma\left[n/2\right]/\Gamma\left[(n+1)/2\right]$.  Notice that Fig.~2 shows not only the collapse of the distribution across stocks but also the normal transport explicitly suggested by Eqs.~(1,3) and observed in Fig.~1.

Finally, in Fig.~3, we focus on the probability of large returns and compare the tail of the observed distribution to that of the predicted distribution for each stock.  For this figure, we measure the slope of the tail of the empirical ccd (in logarithmic coordinates) using the Hill estimator\cite{Hill75} on the largest five percent of the data.  We do this for $\tau=10,20,40,80,160,320$ and average the results (we do not include $\tau=640$ because there are too few data points to get a reliable estimate at this time scale).  This is compared with the slope of the tail from the predicted distribution in the same region.  The measured values are in good agreement with our predictions, showing a pronounced variation across stocks that is explained by our model.  This indicates that the likelihood of extreme price movements is determined by the parameters $\beta_0$ and $n$, obtained from fitting $\beta$ to a gamma distribution for each stock.

\section{Discussion}
We have presented evidence that the non-Gaussian shape and stable scaling of the return distribution are due to slow, but significant, fluctuations in volatility.  Furthermore, our results suggest that return distributions for stocks from different exchanges, time periods, and over different time scales can be described by one functional form.  Because we have only studied well-known stocks from liquid exchanges, it is unknown if this apparent universal behavior for liquid stocks will carry over to stocks that are infrequently traded.

Since the behavior of price fluctuations is rooted in the characteristics of volatility, we expect our results to bring increased interest to stochastic volatility models\cite{Shephard05}, and especially to those that can produce a gamma distributed $\beta$\cite{Nelson90, Bouchaud01, Platen01, PlatenHeath06} (also e-print arXiv:physics/0507073).  Such models can provide important insight into the fundamental mechanism that underlies price fluctuations.

\begin{acknowledgments}
This work was supported by National Science Foundation Grant HSD-0624351.  We thank J. D. Farmer, J. C. Flack, D. C. Krakauer, and J. H. Miller for their helpful comments and suggestions; the Santa Fe Institute for support during this project; and J. A. P\'erez from the Sociedad de Bolsas for data from the SSE.
\end{acknowledgments}


\begin{thebibliography}{24}
\expandafter\ifx\csname natexlab\endcsname\relax\def\natexlab#1{#1}\fi
\expandafter\ifx\csname bibnamefont\endcsname\relax
  \def\bibnamefont#1{#1}\fi
\expandafter\ifx\csname bibfnamefont\endcsname\relax
  \def\bibfnamefont#1{#1}\fi
\expandafter\ifx\csname citenamefont\endcsname\relax
  \def\citenamefont#1{#1}\fi
\expandafter\ifx\csname url\endcsname\relax
  \def\url#1{\texttt{#1}}\fi
\expandafter\ifx\csname urlprefix\endcsname\relax\def\urlprefix{URL }\fi
\providecommand{\bibinfo}[2]{#2}
\providecommand{\eprint}[2][]{\url{#2}}

\bibitem[{\citenamefont{Bachelier}(1900)}]{Bachelier00}
\bibinfo{author}{\bibfnamefont{L.}~\bibnamefont{Bachelier}},
  \bibinfo{journal}{Annales de l'\'{E}cole Normale Sup\'{e}rieure, Ser. 3}
  \textbf{\bibinfo{volume}{17}}, \bibinfo{pages}{21} (\bibinfo{year}{1900}).

\bibitem[{\citenamefont{Osborne}(1959)}]{Osborne59}
\bibinfo{author}{\bibfnamefont{M.~F.~M.} \bibnamefont{Osborne}},
  \bibinfo{journal}{Operations Research} \textbf{\bibinfo{volume}{7}},
  \bibinfo{pages}{145} (\bibinfo{year}{1959}).

\bibitem[{\citenamefont{Mandelbrot}(1963)}]{Mandelbrot63}
\bibinfo{author}{\bibfnamefont{B.}~\bibnamefont{Mandelbrot}},
  \bibinfo{journal}{J. Business} \textbf{\bibinfo{volume}{36}},
  \bibinfo{pages}{394} (\bibinfo{year}{1963}).

\bibitem[{\citenamefont{Fama}(1965)}]{Fama65}
\bibinfo{author}{\bibfnamefont{E.~F.} \bibnamefont{Fama}}, \bibinfo{journal}{J.
  Business} \textbf{\bibinfo{volume}{38}}, \bibinfo{pages}{34}
  (\bibinfo{year}{1965}).

\bibitem[{\citenamefont{Mantegna and Stanley}(1995)}]{Mantegna95}
\bibinfo{author}{\bibfnamefont{R.~N.} \bibnamefont{Mantegna}} \bibnamefont{and}
  \bibinfo{author}{\bibfnamefont{H.~E.} \bibnamefont{Stanley}},
  \bibinfo{journal}{Nature} \textbf{\bibinfo{volume}{376}}, \bibinfo{pages}{46}
  (\bibinfo{year}{1995}).

\bibitem[{\citenamefont{Bouchaud and Potters}(2003)}]{BouchaudPotters03}
\bibinfo{author}{\bibfnamefont{J.~P.} \bibnamefont{Bouchaud}} \bibnamefont{and}
  \bibinfo{author}{\bibfnamefont{M.}~\bibnamefont{Potters}},
  \emph{\bibinfo{title}{Theory of Financial Risks and Derivative Pricing}}
  (\bibinfo{publisher}{Cambridge Univ. Press}, \bibinfo{address}{Cambridge,
  U.K.}, \bibinfo{year}{2003}), \bibinfo{edition}{2nd} ed.

\bibitem[{\citenamefont{Gabaix et~al.}(2003)\citenamefont{Gabaix, Gopikrishnan,
  Plerou, and Stanley}}]{Gabaix03}
\bibinfo{author}{\bibfnamefont{X.}~\bibnamefont{Gabaix}},
  \bibinfo{author}{\bibfnamefont{P.}~\bibnamefont{Gopikrishnan}},
  \bibinfo{author}{\bibfnamefont{V.}~\bibnamefont{Plerou}}, \bibnamefont{and}
  \bibinfo{author}{\bibfnamefont{H.~E.} \bibnamefont{Stanley}},
  \bibinfo{journal}{Nature} \textbf{\bibinfo{volume}{423}},
  \bibinfo{pages}{267} (\bibinfo{year}{2003}).

\bibitem[{\citenamefont{Viswanathan et~al.}(2003)\citenamefont{Viswanathan,
  Fulco, Lyra, and Serva}}]{Viswanathan03}
\bibinfo{author}{\bibfnamefont{G.~M.} \bibnamefont{Viswanathan}},
  \bibinfo{author}{\bibfnamefont{U.~L.} \bibnamefont{Fulco}},
  \bibinfo{author}{\bibfnamefont{M.~L.} \bibnamefont{Lyra}}, \bibnamefont{and}
  \bibinfo{author}{\bibfnamefont{M.}~\bibnamefont{Serva}},
  \bibinfo{journal}{Physica A} \textbf{\bibinfo{volume}{329}},
  \bibinfo{pages}{273} (\bibinfo{year}{2003}).

\bibitem[{\citenamefont{Farmer et~al.}(2004)\citenamefont{Farmer, Gillemot,
  Lillo, Mike, and Sen}}]{Farmer04}
\bibinfo{author}{\bibfnamefont{J.~D.} \bibnamefont{Farmer}},
  \bibinfo{author}{\bibfnamefont{L.}~\bibnamefont{Gillemot}},
  \bibinfo{author}{\bibfnamefont{F.}~\bibnamefont{Lillo}},
  \bibinfo{author}{\bibfnamefont{S.}~\bibnamefont{Mike}}, \bibnamefont{and}
  \bibinfo{author}{\bibfnamefont{A.}~\bibnamefont{Sen}},
  \bibinfo{journal}{Quant. Finance} \textbf{\bibinfo{volume}{4}},
  \bibinfo{pages}{383} (\bibinfo{year}{2004}).

\bibitem[{\citenamefont{Bassler et~al.}(2007)\citenamefont{Bassler, McCauley,
  and Gunaratne}}]{Bassler07}
\bibinfo{author}{\bibfnamefont{K.~E.} \bibnamefont{Bassler}},
  \bibinfo{author}{\bibfnamefont{J.~L.} \bibnamefont{McCauley}},
  \bibnamefont{and} \bibinfo{author}{\bibfnamefont{G.~H.}
  \bibnamefont{Gunaratne}}, \bibinfo{journal}{Proc. Natl. Acad. Sci. U.S.A.}
  \textbf{\bibinfo{volume}{104}}, \bibinfo{pages}{17287}
  (\bibinfo{year}{2007}).

\bibitem[{\citenamefont{Stanley}(2003)}]{Stanley03}
\bibinfo{author}{\bibfnamefont{H.~E.} \bibnamefont{Stanley}},
  \bibinfo{journal}{Physica A} \textbf{\bibinfo{volume}{318}},
  \bibinfo{pages}{279} (\bibinfo{year}{2003}).

\bibitem[{\citenamefont{Press}(1967)}]{Press67}
\bibinfo{author}{\bibfnamefont{S.~J.} \bibnamefont{Press}},
  \bibinfo{journal}{J. Business} \textbf{\bibinfo{volume}{40}},
  \bibinfo{pages}{317} (\bibinfo{year}{1967}).

\bibitem[{\citenamefont{Praetz}(1972)}]{Praetz72}
\bibinfo{author}{\bibfnamefont{P.~D.} \bibnamefont{Praetz}},
  \bibinfo{journal}{J. Business} \textbf{\bibinfo{volume}{45}},
  \bibinfo{pages}{49} (\bibinfo{year}{1972}).

\bibitem[{\citenamefont{Clark}(1973)}]{Clark73}
\bibinfo{author}{\bibfnamefont{P.~K.} \bibnamefont{Clark}},
  \bibinfo{journal}{Econometrica} \textbf{\bibinfo{volume}{41}},
  \bibinfo{pages}{135} (\bibinfo{year}{1973}).

\bibitem[{\citenamefont{Blattberg and Gonedes}(1974)}]{Blattberg74}
\bibinfo{author}{\bibfnamefont{R.~C.} \bibnamefont{Blattberg}}
  \bibnamefont{and} \bibinfo{author}{\bibfnamefont{N.~J.}
  \bibnamefont{Gonedes}}, \bibinfo{journal}{J. Business}
  \textbf{\bibinfo{volume}{47}}, \bibinfo{pages}{244} (\bibinfo{year}{1974}).

\bibitem[{\citenamefont{Lux}(1996)}]{Lux96}
\bibinfo{author}{\bibfnamefont{T.}~\bibnamefont{Lux}},
  \bibinfo{journal}{Applied Financial Economics} \textbf{\bibinfo{volume}{6}},
  \bibinfo{pages}{463} (\bibinfo{year}{1996}).

\bibitem[{\citenamefont{Gardiner}(2004)}]{Gardiner04}
\bibinfo{author}{\bibfnamefont{C.~W.} \bibnamefont{Gardiner}},
  \emph{\bibinfo{title}{Handbook of Stochastic Methods}}
  (\bibinfo{publisher}{Springer-Verlag}, \bibinfo{address}{Berlin},
  \bibinfo{year}{2004}), \bibinfo{edition}{3rd} ed.

\bibitem[{\citenamefont{Andersen and Bollerslev}(1997)}]{Andersen97}
\bibinfo{author}{\bibfnamefont{T.~G.} \bibnamefont{Andersen}} \bibnamefont{and}
  \bibinfo{author}{\bibfnamefont{T.}~\bibnamefont{Bollerslev}},
  \bibinfo{journal}{J. of Empirical Finance} \textbf{\bibinfo{volume}{4}},
  \bibinfo{pages}{115} (\bibinfo{year}{1997}).

\bibitem[{\citenamefont{Bouchaud}(2001)}]{Bouchaud01}
\bibinfo{author}{\bibfnamefont{J.~P.} \bibnamefont{Bouchaud}},
  \bibinfo{journal}{Quant. Finance} \textbf{\bibinfo{volume}{1}},
  \bibinfo{pages}{105} (\bibinfo{year}{2001}).

\bibitem[{\citenamefont{Platen and Heath}(2006)}]{PlatenHeath06}
\bibinfo{author}{\bibfnamefont{E.}~\bibnamefont{Platen}} \bibnamefont{and}
  \bibinfo{author}{\bibfnamefont{D.}~\bibnamefont{Heath}},
  \emph{\bibinfo{title}{A Benchmark Approach to Quantitative Finance}}
  (\bibinfo{publisher}{Springer-Verlag}, \bibinfo{address}{Berlin},
  \bibinfo{year}{2006}).

\bibitem[{\citenamefont{Hill}(1975)}]{Hill75}
\bibinfo{author}{\bibfnamefont{B.~M.} \bibnamefont{Hill}},
  \bibinfo{journal}{The Annals of Statistics} \textbf{\bibinfo{volume}{3}},
  \bibinfo{pages}{1163} (\bibinfo{year}{1975}).

\bibitem[{\citenamefont{Shephard}(2005)}]{Shephard05}
\bibinfo{author}{\bibfnamefont{N.}~\bibnamefont{Shephard}},
  \emph{\bibinfo{title}{Stochastic Volatility: Selected Readings}}
  (\bibinfo{publisher}{Oxford Univ. Press}, \bibinfo{address}{USA},
  \bibinfo{year}{2005}).

\bibitem[{\citenamefont{Nelson}(1990)}]{Nelson90}
\bibinfo{author}{\bibfnamefont{D.~B.} \bibnamefont{Nelson}},
  \bibinfo{journal}{J. Econometrics} \textbf{\bibinfo{volume}{45}},
  \bibinfo{pages}{7} (\bibinfo{year}{1990}).

\bibitem[{\citenamefont{\mbox{E. Platen in:}}(2001)}]{Platen01}
\bibinfo{author}{\bibnamefont{\mbox{E. Platen in:}}},
  \emph{\bibinfo{title}{Mathematical Finance}}
  (\bibinfo{publisher}{Birkhauser}, \bibinfo{address}{Basel},
  \bibinfo{year}{2001}).

\end{thebibliography}
\end{document}